\documentclass[twocolumn,prl,superscriptaddress]{revtex4}

\usepackage{times}
\usepackage{amsmath}
\usepackage{amssymb}
\usepackage{graphicx}
\newcommand{\bra}[1]{\left\langle #1\right|}
\newcommand{\ket}[1]{\left|#1\right\rangle}

\begin{document}

\title{Polynomial-time quantum algorithm for the simulation of chemical
dynamics}

\author{Ivan Kassal}
\affiliation{Department of Chemistry and Chemical Biology, Harvard University, Cambridge, MA 02138}
\author{Stephen P. Jordan}
\affiliation{Department of Physics, Massachusetts Institute of Technology, Cambridge, MA 02139}
\author{Peter J. Love}
\affiliation{Department of Physics, Haverford College, Haverford, PA 19041}
\author{Masoud Mohseni}
\affiliation{Department of Chemistry and Chemical Biology, Harvard University, Cambridge, MA 02138}
\author{Al\'an Aspuru-Guzik}
\affiliation{Department of Chemistry and Chemical Biology, Harvard University, Cambridge, MA 02138}
\email{aspuru@chemistry.harvard.edu}

\begin{abstract}

The computational cost of exact methods for quantum
simulation using classical computers grows exponentially with system
size. As a consequence, these techniques can only be applied to
small systems. By contrast, we demonstrate that quantum computers
could exactly simulate chemical reactions in polynomial time. Our
algorithm uses the split-operator approach and explicitly
simulates all electron-nuclear and inter-electronic interactions in quadratic time.
Surprisingly, this treatment is not only more accurate than the
Born-Oppenheimer approximation, but faster and more efficient as
well, for all reactions with more than about four atoms. This is the case
even though the entire electronic wavefunction is propagated on a
grid with appropriately short timesteps. Although the preparation
and measurement of arbitrary states on a quantum computer is
inefficient, here we demonstrate how to prepare states of chemical
interest efficiently. We also show how to efficiently obtain
chemically relevant observables, such as state-to-state transition
probabilities and thermal reaction rates. Quantum computers
using these techniques could outperform current classical computers
with one hundred qubits.

\end{abstract}

\maketitle

Accurate simulations of quantum-mechanical processes have greatly
expanded our understanding of the fundamentals of chemical reaction dynamics.
In particular, recent years have seen tremendous progress in methods development,
which has enabled simulations of increasingly complex quantum systems. While
it is strictly speaking true that exact quantum simulation requires resources
that scale exponentially with system size, several techniques are available that
can treat realistic chemical problems, at a given accuracy, with only a polynomial
cost. Certain fully quantum methods---such as
multiconfigurational time-dependent Hartree (MCTDH)~\cite{manthe_wave-packet_1992},
matching pursuit/split-operator Fourier transform (MP/SOFT)~\cite{batista_2003},
or full multiple spawning (FMS)~\cite{ben-nun_1998}---solve
the nuclear Schr\"odinger equation, including nonadiabatic effects, given
analytic expressions for the potential energy surfaces and the couplings between them.
These techniques have permitted the simulation of large systems; as examples we can
give MCTDH simulations of a penta-atomic chemical reaction~\cite{wu_first-principles_2004}
and of a spin-boson model with 80 degrees of freedom~\cite{wang_2000}, or an MP/SOFT
simulation of photoisomerization in rhodopsin using 25 degrees of freedom~\cite{batista_2007}.
\textit{Ab initio} molecular dynamics techniques such as \textit{ab initio}
multiple spawning (AIMS)~\cite{ben-nun_ab_2000} avoid analytic expresions for potential energy surfaces
and instead solve electronic Schr\"odinger equation at every timestep. This allows one to gain insight into dynamical problems such as isomerizations through conical
intersections~\cite{benjamin_g_levine_isomerization_2007}.

However, there are also chemical processes which are best treated by
completely avoiding the Born-Oppenheimer approximation. As examples
we can cite strong-field electronic dynamics in atoms and multi-electron
ionization~\cite{parker_double-electron_2001,ruiz_lithium_2005}, or
atomic and molecular fragmentation caused by collisions with
energetic electrons or
photons~\cite{Rescigno12241999,WimVanroose12162005}. Systems
that resist the application of the Born-Oppenheimer approximation
require very general techniques, and the consequent unfavorable
scaling has restricted such simulations to systems with a few particles.
Here, however, we show that the Born-Oppenheimer approximation would
not necessarily simplify simulations performed on quantum computers.
Indeed, except for the smallest systems, an explicit treatment of all the
particles would be both more accurate and more efficient, even for
nearly adiabatic chemical reactions.

Feynman's idea of using a quantum machine to mimic the quantum Hamiltonian of a
system of interest was one of the founding ideas of the field of quantum
computation~\cite{feynman_simulating_1982}. Lloyd~\cite{lloyd_universal_1996} subsequently showed that
quantum computers could be used to simulate systems which can be formulated in
terms of local interactions, using resources
that scale only polynomially with system size. Zalka
and Wiesner~\cite{zalka_simulating_1998,wiesner_simulations_1996} developed a quantum simulation algorithm for particles in real
space and Lidar and Wang~\cite{lidar_calculating_1999}
applied it to the calculation of the thermal rate constant of chemical
reactions. Smirnov \textit{et al.}~\cite{smirnov} proposed an analog
quantum simulator for chemical reactions using quantum dots. We have previously
shown~\cite{aspuru-guzik_simulated_2005} that quantum computers could be used to
simulate the \textit{static} properties of molecules, and in this work we
present a general scheme for using quantum computers for the study of
\textit{dynamical} chemical properties.

To simulate a quantum system we must prepare its initial quantum state,
propagate it in time, and finally extract data of chemical relevance, such as
rate constants. For an efficient quantum simulation, all these tasks must be
carried out using resources which increase polynomially with increasing system
size. We present a quantum algorithm that meets these requirements. We also show
that for all chemical reactions with more than about four atoms, it is more efficient
for a quantum computer to simulate the complete nuclear and electronic
time-evolution rather than to use the Born-Oppenheimer approximation.

The polynomial scaling of these methods means they would enable the study of systems
which are in principle out of reach for any classical computer.
However, large quantum computers are far in the future, and so determining the requirements
of interesting calculations in absolute terms is, perhaps, of more interest than
their scaling alone. We show that a quantum computer using these techniques could
outperform current classical computers using one hundred qubits, within the
design limits of a proposed 300-qubit quantum computer~\cite{steane_to_2004}.
While we focus on chemical applications, these
techniques are generally applicable to many physical systems, from strong-field,
multielectron ionization to quantum liquids and condensed matter systems.

This article is organized as follows. We first review Zalka and Wiesner's
algorithm and show
how the difficulty of computing a wavefunction's time evolution depends only on
the complexity of evaluating the interaction potential. We then consider three
approaches to the calculation of the interaction potential, including a fully
non-adiabatic treatment of chemical reactions. We consider the problem of state
preparation for all of the schemes, and finally address the problem of
measurement. We present three readout schemes for reaction dynamics---reaction
probabilities, thermal rate constants, and state-to-state probabilities---which
would allow for efficient evaluation of many parameters accessible to
experiment.

\section{Quantum Dynamics}

The problem of simulating quantum dynamics is that of determining the properties
of the wavefunction $\left|\psi(t)\right\rangle $ of a system at time $t$, given
the initial wavefunction $\left|\psi(0)\right\rangle$ and the Hamiltonian
$\hat{H}$ of the system. If the final state can be prepared by propagating the
initial state, any observable of interest may be computed.

We employ an improved version of the real-space quantum simulation technique
developed by Zalka and Wiesner in which a discrete variable representation of
the wavefunction is used~\cite{zalka_simulating_1998,wiesner_simulations_1996}.
In the one-dimensional case, the domain of the wavefunction is divided into a
discrete position basis of $N=2^{n}$ equidistant points. The wavefunction is
represented as:
\[
\ket{\psi(t)} = \sum_{x=0}^{2^{n}-1}a_{x}(t)|x\rangle
= a_{0}\underbrace{\left|0\ldots00\right\rangle }_{n\:\mathrm{qubits}}+
\ldots+a_{2^{n}-1}\left|1\ldots11\right\rangle .
\]
The spatial wavefunction is stored in the Hilbert space of the qubits, and so
the spatial resolution grows exponentially
with the number of qubits. For a system with $d$ dimensions, $d$ registers of
$n$ qubits each are used,
$\ket{\mathbf{x}}=\ket{x_1}\cdots\ket{x_d}$, representing a grid of $2^{dn}$
points.  The states
of multiple particles can be stored by adding position registers for each
particle. Therefore, only
a polynomially large number of qubits is required to store the system
wavefunction.

For simplicity we assume a time-independent Hamiltonian whose potential depends
only on position,
$\hat{H}=\hat{T}+\hat{V}$ where $\hat{T} = \hat{p}^2/2m$ and $\hat
{V}=V(\hat{\mathbf{x}})$ are the kinetic and potential energy operators,
respectively. The split operator
method~\cite{feit_solution_1982,kosloff_fourier_1983,zalka_simulating_1998}
computes the time evolution by separating the kinetic $\hat{T}$ and potential
$\hat{V}$ energy contributions to the propagator $\hat{U}(t)=e^{-i\hat{H}t}$.
Given a sufficiently small time step $\delta t$, we can write to first order
\begin{equation*}
\hat{U}(\delta t)=e^{-i\hat{H}\delta t}= e^{-i\hat{T}(\mathbf{x})\delta
t}e^{-i\hat{V}(\mathbf{x})\delta t} + O(\delta t^2).
\end{equation*}
The operators $e^{-i\hat{V}\delta t}$ and $e^{-i\hat{T}\delta t}$ are
diagonal in the position and momentum representations, respectively. A
quantum computer can efficiently transform between the two
representations using the quantum Fourier transform (QFT)
\cite{coppersmith_approximate_2002}:
\begin{equation*}
\left|\psi(\delta t)\right\rangle =\hat{U}(\delta t)\left|\psi(0)\right\rangle
\approx \text{QFT}\, e^{-iT(\mathbf{p})\delta
t}\mathrm{\,\text{QFT}^{\dagger}\,}e^{-iV(\mathbf{x})\delta
t}\left|\psi(0)\right\rangle.
\end{equation*}
The procedure is iterated as many times as necessary to obtain the system
wavefunction
$\ket{\psi(t)}$ after an arbitrary time $t$ to a desired accuracy.

\begin{center}
\begin{figure}
\includegraphics[width=3in]{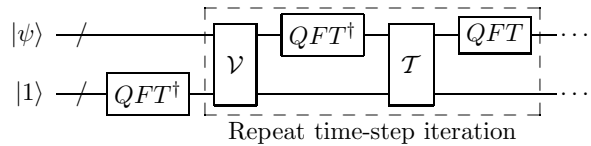}
\caption{The quantum simulation algorithm. The potential and kinetic energy
  unitaries are applied to a quantum state in turn, with the transformation
between
  position and momentum representations being performed with the efficient
  quantum Fourier transform (QFT). The ancilla register is required for phase
  kickback and remains unchanged throughout the simulation, while the boxed time
  step is repeated $t/\delta t$ times. The proposed algorithm, unlike that of
  Zalka~\cite{zalka_simulating_1998}, does not require that functions be
  uncomputed and is therefore twice as fast. \label{fig:circuit}}
\end{figure}
\end{center}

The application of diagonal unitaries is straightforward on a quantum computer.
Suppose that we have a gate sequence which acts on an arbitrary position
eigenstate as $\ket{\mathbf{x}} \rightarrow e^{-iV(\mathbf{x})\delta
  t}\ket{\mathbf{x}}$. Since $\ket{\psi}$ is a superposition of position
eigenstates, when this gate sequence is applied to $\ket{\psi}$, one obtains
$e^{-i\hat{V}\delta t}\ket{\psi}$ in a single application.

We depart from Zalka and Wiesner's method in the implementation of this gate
sequence. We are free to take the lowest value of the potential
in the domain as $0$, and use such units that the maximum value of the potential
is $V_{\rm max}=2^m-1$, with $m$ an integer. With this choice of units $V$ takes
integer values, and we choose $m$ large enough that $V$ is resolved with
sufficient precision. The integer $m$ is therefore the number of qubits required to
represent the desired range of potential values with the desired precision.
The gate sequence $\mathcal{V}$ which computes the potential $V$ acts so that
$\mathcal{V}\left|\mathbf{x},y\right\rangle
=\left|\mathbf{x},y\oplus V(\mathbf{x})\right\rangle $, where
$y$ is an $m$-bit integer labeling a basis state of the ancilla register  and
$\oplus$ denotes addition modulo $2^{m}$.

We apply the diagonal unitary by phase kickback. The computer is
initialized in the state $\left|\psi\right\rangle\otimes\left|1\right\rangle_{m} $, where
$\left|1\right\rangle_{m}$ in the ancilla register represents the state
$\left|0\ldots001\right\rangle$ in $m$ qubits. Applying the
inverse QFT to the ancilla register, followed by $\mathcal{V}$, produces
\begin{equation*}
\mathcal{V}\left(|\psi\rangle \otimes\sum_{y=0}^{M-1}\frac{e^{2\pi i\,y/M}}{\sqrt{M}}\left|y\right\rangle \right)
 = e^{-i\hat{V}\delta t}\left|\psi\right\rangle \otimes\sum_{y=0}^{M-1}\frac{e^{2\pi i\,y/M}}{\sqrt{M}}\left|y\right\rangle 
\end{equation*}
where $M=2^{m}$ and we choose $\delta t=\frac{2\pi}{M}$. The equality
obtains since the ancilla state is an eigenstate
of addition (with eigenvalue $e^{-2\pi iq/M}$ corresponding to the addition of
$q$) \cite{bernstein_quantum_1993}. We see that applying  $\mathcal{V}$ results
in  the requisite diagonal unitary action on the wavefunction register. The
states of register and ancilla are separable before and after each potential
evaluation. We can also define a quantum gate sequence $\mathcal{T}$ which
computes the kinetic energy $p^2/2m$:
$\mathcal{T}\ket{\mathbf{p},y}=\ket{\mathbf{p},y\oplus T(\mathbf{p})}$. This
gate is diagonal in the momentum basis, and has efficiently computable entries
on the diagonal (namely $p^2$). Thus, we use the quantum Fourier transform to
conjugate into the momentum basis  and $\mathcal{T}$ is implemented by phase
kickback in exactly the same way as $\mathcal{V}$. The quantum circuit for this
algorithm is shown in Fig.~\ref{fig:circuit}.

This simulation algorithm is numerically exact in the sense that all introduced
approximations are controlled, so that the error in the calcuation can be arbitrarily 
reduced with an additional polynomial cost. The only approximations employed are the
discretization of time, space, and the potential $V(\mathbf{x})$.
The error due to discretization can be made exponentially small by adding more qubits. The error
due to time discretization can be systematically improved by use of higher-order
Trotter schemes~\cite{BerryAhokas06}. The computational cost of the algorithm
per iteration is the evaluation of $V(\mathbf{x})$, $T(\mathbf{p})$ and two
QFT's. While the QFT's and the quadratic form in the kinetic energy ($p^2$ in
the simplest case) can be computed in polynomial
time~\cite{coppersmith_approximate_2002,knuth_art_1997}, the evaluation of the
potential energy $V(\mathbf{x})$ may not be
efficient in general. For example, a random potential stored in an exponentially
large database requires an exponentially
large number of queries to evaluate. However, any classical algorithm
running in $O(f(n))$ time can be adapted to a reversible quantum algorithm also
running in $O(f(n))$ time \cite{nielsen_quantum_2000}. Therefore, the potential
energy $V(\mathbf{x})$ will be efficiently calculable on a quantum computer if
it is efficiently calculable on a classical computer. Fortunately, this is true
for all chemically relevant cases.

\section{Chemical Dynamics}

Every isolated system of chemical interest has the same Hamiltonian,
which in atomic units is
\begin{equation*}
\hat{H}=\sum_i \frac{p_i^2}{2M_i} + \sum_{i<j} \frac{q_iq_j}{r_{ij}},
\end{equation*}
where the sums are over the nuclei and electrons, $p_i$ is the
momentum of the $i^{\mathrm{th}}$ particle, $M_i$ its mass, $q_i$ its
charge and $r_{ij}$ is the distance between particles $i$ and
$j$. Both the potential and kinetic terms can be efficiently evaluated
since the arithmetical operations can be performed in $O(m^2)$ time
\cite{knuth_art_1997}, and for a system of $B$ particles, there are
only $O(B^2)$ terms in the sum.
\footnote{Other $N$-body potentials could also be
efficiently evaluated, with cost of $O\left(m^2B^N\right)$ if they only contain
arithmetic that requires $O(m^2)$ time. For example, the Lennard-Jones
potential could be computed using $\left(\frac{75}{4}m^3+\frac{51}{2}m^2\right)$
gates per pair of particles.
Simulating potentials other that the Coulomb potential could be applied to
situations such as liquid helium clusters, and although we do not discuss
them in detail, the present algorithm could simulate such potentials with
minimal adjustments.}

The fact that the Coulomb potential can be evaluated in $O(B^2m^2)$ time implies
that chemical dynamics could be simulated on a quantum computer in $O(B^2m^2)$
time, an exponential advantage over known classical techniques for exact quantum
simulation. Here $B$ is the number of particles and $m$ is the binary precision
the potential in the region of interest. We want to emphasize
that a quantum simulation would be substantially different from what is usually
done on classical computers. Most significantly, we are proposing to explicitly
track all the nuclei and electrons on a Cartesian grid which is sufficiently
fine and with time steps sufficiently short to capture the true electronic
dynamics. We will show that this is not only more accurate, but also requires
fewer quantum resources.

\begin{center}
\begin{figure}
\includegraphics[width=3in]{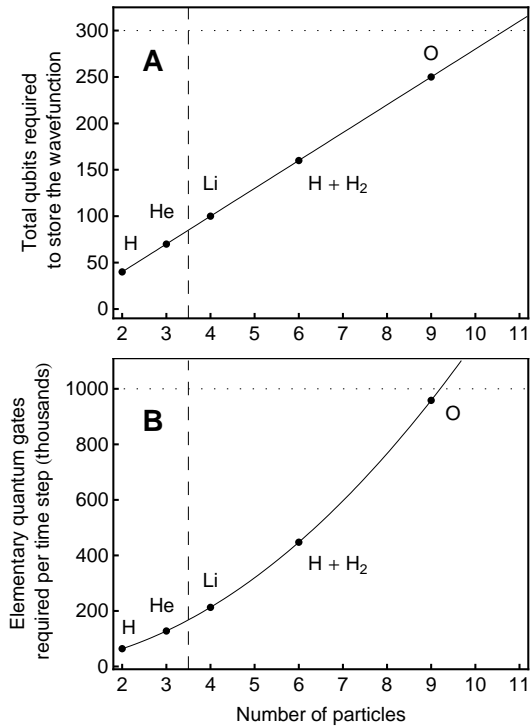}
\caption{Resource requirements for a quantum simulation of $B$ particles
interacting
  through a pairwise potential. The chemical symbols correspond to the
  simulation of the full Coulomb dynamics of the corresponding atom (nucleus and
  electrons). The vertical dashed line represents the approximate current limit
  of numerically exact quantum simulation on classical computers on a grid~\cite{ruiz_lithium_2005}.
  (\textit{A}) Total qubits required. We require $n$ qubits for each degree of
  freedom and $m$ qubits for each ancilla, four of which are needed for the Coulomb potential.
  Hence, a total of $n(3B-6)+4m$ qubits are needed (see Supplementary Information
  for details). The horizontal
  dotted line represents a 300-qubit quantum computer, which is believed to be
  feasible with near-future technology \cite{steane_to_2004}. We assume a grid of
  $2^{30}$ points, which corresponds to $n=10$ and would suffice for the
  simulation of many chemical reactions or the strong-field ionization of
  atoms~\cite{schneider_parallel_2006, parker_double-electron_2001}.
  (\textit{B}) Total elementary gates required. The 300-qubit computer is
  expected to achieve one billion elementary quantum operations. The dotted line
  represents the largest possible simulation of $1000$ time steps, assuming ten
  bits of numerical accuracy $(m=10)$. Computing the Coulomb potential
  requires $\left(\frac{75}{4}m^3+\frac{51}{2}m^2\right)$ gates
  per pair of particles (see Supplementary Information for details).
\label{fig:resources}}
\end{figure}
\end{center}

The Supplementary Information contains a detailed computation of the numbers of
gates and qubits required for the quantum simulation of the Coulomb potential.
The number of elementary gates required to evaluate this potential in three
dimensions is $(\frac{75}{4}m^3+\frac{51}{2}m^2)$ per pair of particles (Fig.~\ref{fig:resources}). We
chose a method which minimizes the number of ancilla qubits and so is suited
for small numbers of qubits. Note that this scaling is not asymptotically
optimal (the asymptotic requirement would be $O(m^2)$), so further improvement could
be achieved for computations with high precision (large $m$) if suitable
arithmetical algorithms were implemented.
Storing the wavefunction of a system with $d$ degrees of
freedom requires $nd$ qubits, so a system of $B$ particles, with $d=3B-6$
degrees of freedom, requires $n(3B-6)$ qubits. To this one must add the qubits
needed for the ancilla registers, only four of which are required for the
Coulomb potential, meaning that simulating these potentials requires $n(3B-6)+4m$
qubits (Fig.~\ref{fig:resources}).

On a small quantum computer, the computational cost of simulating the
interactions between many particles may prove prohibitive. One could try to
simplify matters and focus only on the nuclear motion by employing the Born-Oppenheimer
approximation. Here, the solution of the electronic structure problem provides a
potential energy surface on which the nuclei move. However, we show that quantum computers
would benefit from the Born-Oppenheimer approximation only rarely: for many chemical
reactions, simulating the full dynamics of both electrons and nuclei will not
only yield more accurate results, but will also, remarkably, be faster. This is
in sharp contrast to the study of chemical dynamics on classical computers,
where there is frequent need to simplify calculations using the Born-Oppenheimer
approximation.

\begin{center}
\begin{figure}
\includegraphics[width=2.5in]{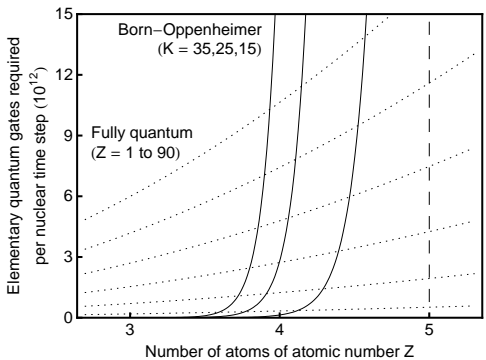}
\caption{Estimated number of elementary quantum operations (gates)
  required for the simulation of
  chemical reactions. Standard Born-Oppenheimer potential-energy-surface
  calculations require time resources exponential in the size of the system
  (full line), while an fully nonadiabatic, nuclear and electronic calculation
  would require only polynomial time (dotted). The resulting cutoff indicates
  that for all reactions with more than four atoms (dashed) the
  Born-Oppenheimer approximation is always less efficient on a quantum computer
  than a diabatic treatment. The complexity of the diabatic computation depends
  only on the atomic number $Z$, while the potential energy surfaces are modeled
  as polynomials of degree $K$ along each axis. A value of $K\ge15$ is
  required to obtain 0.1\% agreement with surfaces such as
  BKMP2~\cite{boothroyd_refined_1996}. The position of the cutoff
  does not significantly depend on the accuracy of the evaluated potential
  ($m$). To obtain the gate counts, we assume 20 bits of accuracy $(m=20)$,
  enough for chemical precision. The gate counts reflect the fact that an appropriate
  nuclear time step is about 1000 times longer than an electronic time
  step. \label{fig:cutoff}}
\end{figure}
\end{center}

It is difficult to estimate the precise computational cost of using the
Born-Oppenheimer approximation, since different potential energy surfaces have
different functional forms. Nevertheless, for any general fitting technique, the
complexity of the interpolating function grows exponentially with increasing
dimension of the system, since exponentially many data points must be used in
the fit if one is to maintain uniform accuracy over the entire domain. We can
provide an estimate of the computational cost for a potential energy surface
which is an interpolating polynomial of degree $K$ along each dimension (see
Supplementary Information). In that case, the total cost of the adiabatic
simulation is $K^{3B-6}\left(\frac54m^3+\frac52m^2\right)$ per nuclear time step (which is
usually about a thousand times longer than an electronic time step). Numerical
experiments with the BKMP2 surface for
$\mathrm{H_3}$~\cite{boothroyd_refined_1996} indicate that $K$ must be chosen to
equal at least 15 if one aspires to 0.1\% accuracy in the potential, and more
for chemical accuracy. With $K=15$, the exponential growth implies that even for
heavy elements ($Z\approx 100$), the fully-dimensional diabatic treatment is
faster for all reactions involving more than four atoms, and even for many smaller
reactions, as shown in Fig.~\ref{fig:cutoff}.

It is perhaps beneficial to briefly discuss the intuitive reasons why the use of
pre-computed potential energy surfaces
is not as useful on quantum computers as it is on classical machines.
Classically, an exponential amount of memory is required in general to store the
wavefunction. However, the ratio of computing power to memory in most classical
computers is large and the basic floating point operations are hardwired. Since
the storage capacity is often the limiting factor, if a wavefunction can be stored
in memory, its motion on a surface can probably be computed.
Quantum computers, on the
other hand, require only linearly many qubits to store the wavefunction in a
superposition state. However, using a pre-computed potential requires either
the evaluation of a complicated function or a look-up in a potentially
large table. The potential energies must be computed on the fly in
order to take advantage of quantum parallelism and it is therefore imperative to
keep the interaction potential as simple as possible. This is achieved by treating all the
particles explicitly, interacting via the Coulomb interaction.

An alternative way to compute a potential energy surface would be to embed an
on-the-fly calculation of electronic structure in the quantum algorithm and thus
avoid a classically precomputed fit. This can be done efficiently since
electronic structure calculations can be performed in polynomial time on quantum
computers~\cite{aspuru-guzik_simulated_2005}. Hence, the quantum circuit
$\mathcal{V}$ would be replaced by a black box containing the efficient quantum
version of the full configuration interaction (FCI) procedure
\cite{aspuru-guzik_simulated_2005}. Because the quantum simulation algorithm
exploits quantum effects, a single evaluation of electronic structure is
sufficient for each time step: all the nuclear configurations are evaluated in
superposition. However, the electronic structure circuit for the proposed
algorithm would require the atomic positions as input. This would require a
modification of the original algorithm so that the Coulomb and exchange
integrals are computed using a quantum circuit rather than classically. While
this approach, unlike the Born-Oppenheimer approximation, is asymptotically
appealing, the large overhead required to compute the exchange integrals quantum
mechanically makes it uninteresting for near-future implementation.

Steane has recently proposed a design for a 300-qubit, error-corrected,
trapped-ion quantum computer that could perform around $10^9$ quantum operations
using methods for quantum gates that have already been experimentally
implemented~\cite{steane_to_2004}. On a three-dimensional grid of $2^{30}$
points, such a computer could store the wavefunction of a ten-particle system
(Fig. \ref{fig:resources}). By comparison, classical computers implementing a
comparable grid based algorithm are limited to computing the full quantum
evolution of a three-particle system, such as a helium
atom~\cite{parker_double-electron_2001,ruiz_lithium_2005}. Even a relatively
modest quantum computer with 100 qubits could simulate the electron dynamics or
ionization of the lithium atom, a task beyond the reach of classical computers
using grid based methods~\cite{ruiz_lithium_2005}.  The simplest chemical
reaction, $\mathrm{H+H_2\rightarrow H_2+H}$, is a six-particle system, and could
therefore be simulated by Steane's computer in a fully-dimensional diabatic
regime. While other classical methods may be able to reach somewhat larger
examples, the exponential scaling of all known classical exact methods means
that the examples given here are close to the crossover point between classical
and quantum computing for chemical dynamics. There remain two questions: how to
prepare the initial state of the quantum computer, and how to extract useful
information out of the final state.

\section{State Preparation}

The preparation of an arbitrary quantum state is exponentially hard in
general~\cite{SZK}.
Nevertheless, we show that most commonly used chemical wavefunctions can be
prepared efficiently. Since the significant deviations from Born-Oppenheimer
behavior occur during evolution and usually do not concern initial
states, we will prepare the initial state using the
Born-Oppenheimer approximation. That is, the system wavefunction
will be a product state
$\ket{\psi}=\ket{\psi_{\mathrm{nuc}}}\ket{\psi_{\mathrm{elec}}}$
of nuclear and electronic wavefunctions, each in its own
quantum register.

Nuclear motions can be expressed in normal mode coordinates if the
displacements from equilibrium are small, which is the case in
molecules at chemically relevant temperatures. The nuclear
wavefunction is then, along each coordinate, a superposition of
harmonic oscillator eigenstates, which are themselves products of
Gaussians and Hermite polynomials. It is known that
superpositions corresponding to efficiently integrable functions can
be prepared on a quantum computer in polynomial
time~\cite{zalka_simulating_1998,grover_creating_2002}.
Therefore, Gaussian wavepackets and Hermite polynomials are
efficiently preparable, meaning that we can prepare good
approximations to molecular vibrational states and Gaussian wave
packets.

Gaussian wavepackets can also be used to prepare the electronic
wavefunction. Indeed, it is customary in electronic structure theory
to expand molecular orbitals in terms of atomic orbitals,
which themselves are superpositions of Gaussian-type orbitals.
The orbital occupation numbers can be obtained from
electronic structure calculations, including our earlier quantum
algorithm~\cite{aspuru-guzik_simulated_2005}. Consequently, the occupied
orbitals, which are superpositions of Gaussians, can all be prepared
efficiently.

One final consideration is the exchange symmetry of multi-electron
wavefunctions. Abrams and Lloyd proposed a method for preparing
antisymmetric wavefunctions~\cite{abrams_simulation_1997} starting with a
Hartree product of molecular orbitals. We propose to use this method for
preparation of multi-electron wavefunctions, noting that it suffices to prepare the
initial state with the correct exchange symmetry, since the exchange
operator commutes with the Hamiltonian.

Of course, other strategies for state preparation can be pursued,
such as the phase estimation algorithm
\cite{abrams_quantum_1999}. If we are able to prepare a state
$|S\rangle$ that has a significant overlap $\langle S|E\rangle$ with
an eigenstate $|E\rangle$ (not necessarily the ground state),
phase estimation followed by measurement
will collapse the wavefunction to the desired eigenstate with
probability $\left|\langle S|E\rangle\right|^2$. Alternatively, the
ground state can be prepared by the adiabatic state preparation
algorithm~\cite{aspuru-guzik_simulated_2005}. This is of particular
significance to the simulation of full chemical dynamics, since the
electronic ground state is usually a good approximation for the
reactants.

\section{Measurement}

After preparing the initial state and simulating its time evolution using the
methods described above, we must extract chemically relevant
information from the final system wavefunction. In general, quantum
tomography is the most general approach to the
estimation of an unknown quantum state or a quantum process
\cite{nielsen_quantum_2000} by measuring the expectation values of a
complete set of observables on an ensemble of identical quantum
systems. However, this full characterization of quantum systems
always requires resources that grow exponentially with the size of
the system. In order to avoid such problems, alternative approaches
for the direct estimation of certain properties of quantum dynamical
systems have been recently
developed~\cite{MohseniLidar06,Emerson07}. Here we likewise show
that the data of greatest chemical significance can be obtained
directly with only a polynomial number of measurements. In
particular, we present algorithms for obtaining the reaction
probability, the rate constant, and state-to-state transition
probabilities.

The reaction probability, given a certain initial wavefunction of the
reactants,
is the likelihood of observing, after a sufficiently long time,
the products of the chemical reaction. To find it, we divide the
real-space domain of the wavefunction into $r$ disjoint regions
corresponding to sub-domains of chemical interest. In chemistry
these regions are typically a few simple polytopes. The simplest
division is into only two regions, one for the reactants and one for
the products, separated by the transition state dividing surface
(TSDS). The reaction probability is the sum of the probabilities of
finding the final wavepacket in the product region(s).
It is straightforward to construct a classical point location circuit for
the function $R(\mathbf{x})$ which, given a nuclear
position vector $\mathrm{x}$, identifies which region it is in by returning an
integer label corresponding to that region.
There is a corresponding reversible circuit that  performs the transformation
$\ket{\mathbf{x}}\ket{y} \to \ket{\mathbf{x}} \ket{y \oplus R(\mathbf{x})}$.
We can apply this circuit to the final state
$\ket{\psi}=\sum_{\mathbf{x}}a_{\mathbf{x}}\ket{\mathbf{x}}$,
to which we add an additional ancilla register with
$\lceil \log_2 r\rceil$ qubits. That is, applying this reversible
circuit to $\ket{\psi} \ket{0}$ yields
$\sum_{\mathbf{x}}a_{\mathbf{x}}\ket{\mathbf{x}}\ket{R(\mathbf{x})}$.
Measuring the ancilla register will return $i$ with probability $P_i$,
which equals the probability of finding the wavepacket in the region
$i$. We can obtain all the probabilities $P_{i}$ by employing an
ensemble measurement of the ancilla register. Since individual
measurements are uncorrelated, the error of the estimate of the
probabilities decreases as $1/\sqrt{M}$ for $M$ repetitions of the
experiment. However, it is possible to achieve a convergence of $1/M$,
which is the ultimate limit of quantum
metrology~\cite{giovannetti_quantum_2006}, using techniques such as
amplitude estimation~\cite{brassard_quantum_2000,knill_optimal_2007}.
Next, we use these disjoint regions to compute the rate constant.

The rate constant $k(T)$ at temperature $T$ is a thermally weighted average of
cumulative reaction probabilities~\cite{tannor}:
\begin{eqnarray*}
k(T)& =       & \frac{1}{2\pi\hbar\, Q(T)}\int_0^{\infty} N(E)\, e^{-E/k_BT} dE\\
    & \approx & \frac{1}{2\pi\hbar\, Q(T)}\sum_{\boldsymbol{\zeta},E} P_r(\boldsymbol{\zeta},E)\, e^{-E/k_BT} \Delta E,
\end{eqnarray*}
where $E$ is the energy, $Q(T)$ is the reactant partition function, and $N(E)$
is the cumulative reaction probability,
$N(E)=\sum_{\boldsymbol{\zeta}}P_r(\boldsymbol{\zeta},E)$. The vector
$\boldsymbol{\zeta}$ is a specification of all the quantum numbers of the
reactants and $P_r(\boldsymbol{\zeta},E)$ is the reaction probability given
that the reactants are in the eigenstate specified by $\boldsymbol{\zeta}$
and $E$. The sum ranges over all possible $\boldsymbol{\zeta}$, and with $E$
from zero to a cutoff. Note that on a quantum computer the cutoff can be 
made exponentially large, or the energy step $\Delta E$ exponentially small,
by adding more qubits.

We can compute the rate constant on a quantum computer if we propagate in
time not a pure wavefunction but the correct thermal mixed state. In that case,
the expectation value of the reaction probability would equal the rate constant,
up to a known factor. The required initial state is
$\rho(0)=C^2\sum_{\boldsymbol{\zeta},E}\Gamma\left(E,T\right)^2
  \ket{\phi_0\left(\boldsymbol{\zeta},E\right)}\bra{\phi_0\left(\boldsymbol{\zeta},E\right)}$
where $\Gamma(E,T)=\left( \exp(-E/k_BT) \Delta E / 2\pi\hbar\,Q(T) \right)^{1/2}$
is the square root of the appropriate Boltzmann factor, $C$ is a normalization
constant, and
$\ket{\phi_0\left(\boldsymbol{\zeta},E\right)}$ is a real-space reactant eigenfunction
corresponding to quantum numbers $\boldsymbol{\zeta}$ and energy $E$.
If we propagate $\rho(0)$ for time $t$ using the simulation algorithm,
the system will be in the final state
$\rho(t)=C^2\sum_{\boldsymbol{\zeta},E}\Gamma\left(E,T\right)^2
  \ket{\phi_t\left(\boldsymbol{\zeta},E\right)}\bra{\phi_t\left(\boldsymbol{\zeta},E\right)}$,
where $\ket{\phi_t\left(\boldsymbol{\zeta},E\right)}$ now denotes the
time-evolved version of state $\ket{\phi_0\left(\boldsymbol{\zeta},E\right)}$
(note that, except in exceptional cases, $\ket{\phi_t\left(\boldsymbol{\zeta},E\right)}$
is not an eigenfunction of either reactants or products).
If we have a quantum register in this mixed state, we can add an ancilla qubit
and use the technique of dividing the domain into reactant and product regions as
described above. Finally, a measurement of the ancilla qubit produces $\ket{1}$
with probability $C^2 k(T)$.
The precision of $k(T)$ thus obtained goes as $1/M$
with the number of measurements if we use amplitude estimation. The previously
proposed approach for estimating reaction rates \cite{lidar_calculating_1999}
evaluates the rate constant by computing a flux-flux correlation function based
on a polynomial-size sample of the wavefunction in the position basis. In
contrast, our approach carries out the integration explicitly on the quantum
computer using amplitude amplification, which provides a quadratic improvement
over algorithms that rely on classical sampling and post-processing.

The thermal state $\rho(0)$ can be prepared efficiently on a quantum computer.
We begin by preparing a superposition of the reactant state labels in terms
of $\boldsymbol{\zeta}$ and $E$, i.e., 
$C\sum_{\boldsymbol{\zeta},E}\Gamma\left(E,T\right)\ket{\boldsymbol{\zeta},E}$,
with $C$ and $\Gamma$ as defined above. Here, $\ket{\boldsymbol{\zeta},E}$
contain only the state labels, and not position-space wavefunctions.
If we assume that the thermally accessible states can be
enumerated by a polynomial number of qubits and that the energy can be
specified to a certain precision $\Delta E$, we see that the states
$\ket{\boldsymbol{\zeta},E}$ require only polynomially many qubits to store.
The superposition itself can be prepared efficiently since
$\Gamma(E,T)$ is an exponential function of the energy and is therefore efficiently
preparable~\cite{zalka_simulating_1998,grover_creating_2002}. 

The next step is to generate, by doing state preparation in superposition, the state
$\ket{\Phi_0}=C\sum_{\boldsymbol{\zeta},E}\Gamma\left(E,T\right)\ket{\boldsymbol
  {\zeta},E}\ket{\phi_0\left(\boldsymbol{\zeta},E\right)}$, in which each term
is the tensor product of $\ket{\boldsymbol{\zeta},E}$ and
the corresponding real-space reactant eigenstate $\ket{\phi_0\left(\boldsymbol{\zeta},E\right)}$.
The states $\ket{\phi_0\left(\boldsymbol{\zeta},E\right)}$ must have definite
energy $E$. Hence, one can represent an initial state as a direct product
of discrete reactant internal states (specified by $\boldsymbol{\zeta}$ with energy
$E(\boldsymbol{\zeta})$) and a wavepacket with kinetic energy $E_k = E-E(\boldsymbol{\zeta})$~\cite{tannor}.
The discrete part can be prepared using the state-preparation approach described
above. The incoming wavepacket can be approximated with a Gaussian with a
kinetic energy expectation value of $E_k$. This approximation can be improved by
increasing the width of the Gaussian, which can be doubled by the
addition of a single qubit to the position register. This would halve the
momentum uncertainty of the wavepacket. With sufficiently many qubits, the error
incurred in this approximation could be made smaller than errors from other sources,
such as grid discretization.
Once we have prepared $\ket{\Phi_0}$, we will no longer use the register containing
the states $\ket{\boldsymbol{\zeta},E}$. If we trace this register out, we can see
that the remaining state register is a mixed state with density operator
$\rho(0)$, as desired.

Finally, we show how to obtain state-to-state transition probabilities. Most
chemical reactions can be regarded as scattering processes, and it is therefore
desirable to know the scattering $S$-matrix. In particular, it is these
state-to-state transition amplitudes which are accessible to experiment.
Heretofore we have considered the joint wavefunction of all the molecular species.
To compute state-to-state probabilities, however, we must ensure that each reactant
molecule is in a well-defined molecular state. For example, to probe the state-to-state
dynamics of the $\mathrm{H+H}_2$ reaction, we would need to prepare a particular state
of the hydrogen atom plus a state of the hydrogen molecule, and not simply a state
of the overall $\mathrm{H}_3$ aggregate. Given these states, prepared in the
center-of-mass coordinate systems of each molecule, one must perform coordinate
transformations to put them on a common grid. For each molecule, the Cartesian
coordinates of the particles are linear, invertible functions of their center-of-mass
coordinates. Since the coordinate transformations can be computed by an efficient,
reversible, classical circuit, they can also be efficiently computed quantum
mechanically.

We concentrate here on obtaining only the vibrational state-to-state
distributions. Using the techniques of state preparation above, we prepare each
reactant molecule in a vibrational eigenstate, so that along each of its normal
mode coordinates the molecule is in an eigenstate of the corresponding potential.
After all the molecules are thus prepared, their wavefunctions are transformed
to a common Cartesian system. This initial state is evolved as usual until the
molecules separate into isolated products.

At large inter-molecular separation, the center-of-mass motion and the normal
mode coordinates again become separable. Therefore, an orthogonal
transformation can be applied to each product molecular fragment so that its Cartesian
coordinates can be transformed into normal modes. The quantum phase estimation
algorithm can then be employed to extract the populations and eigenenergies of the
product vibrational states.

For an isolated product molecule, we can expand the final state in terms of the
normal modes:
$\ket{\Psi'}=\sum_{\mathbf{v'}}\alpha'_{\mathbf{v'}}\ket{\boldsymbol{\xi'}_{
    \mathbf{v'}}}$, where $\ket{\boldsymbol{\xi'}_{\mathbf{v'}}}$ is the
position representation of the eigenstate corresponding to product occupation
number vector $\mathbf{v'}$. The state-to-state transition probabilities are
then $P_{\mathbf{v'}\leftarrow\mathbf{v}}=\left|\alpha'_\mathbf{v'}\right|^2$,
and as mentined above, they can be determined using the phase estimation
algorithm of Abrams and Lloyd \cite{abrams_quantum_1999} for each degree of
freedom. We can obtain good measurement statistics with only a polynomial number of
measurements because at typical temperatures, the products of chemical reactions
will have appreciable population in only a small number of vibrational
eigenstates.

\section{Conclusion}

The advantages of the methods presented here are not limited to chemical
reaction dynamics, but can be applied to many areas of physics.
This is true in particular because the complexity of the algorithm is
proportional to the complexity of calculating the potential and because
the laws of nature are usually captured by simple, few-body
interactions. For example, by using a quantum computer to study atoms in the
presence of a strong,
time-dependent electric field, one could simulate such effects as
multielectron ionization or attosecond pulse
generation~\cite{parker_double-electron_2001,ruiz_lithium_2005,
schneider_parallel_2006}.
Quantum computers also offer the promise of predicting real-time
properties of superfluids~\cite{aziz_examination_1991,ceperley_path_1995},
and of providing tests for effective potentials for water
phases~\cite{fanourgakis_quantitative_2006}.

We close by reiterating the need for a careful reexamination of the
suitability of traditional quantum approximations for use on quantum
computers. Previously we have shown that a quantum implementation of
full configuration interaction scales better than coupled cluster
approaches (in particular CCSD(T)), and in this work we show that
simulating the complete nuclear and electronic time evolution is
more efficient on quantum computers than using the Born-Oppenheimer
approximation, a central tool of theoretical chemistry. We can
imagine the development of a wide variety of new techniques and
approaches tailored for natural quantum simulators, which,
themselves relying on the rules of quantum mechanics, give us a
deeper understanding of physical phenomena.

\begin{acknowledgments}
We wish to thank Eric J. Heller and Jacob Biamonte for valuable discussions,
as well as the Army Research Office (project W911NF-07-1-0304 and the QuaCCR
program) and the Joyce and Zlatko Balokovi\'c Scholarship for funding.
\end{acknowledgments}

\bibliographystyle{apsrev}

\section{Supplementary Information: Quantum Resource Estimation}

There are two kinds of quantum register in the simulation algorithm:
the state registers, used to store the wavefunction, and the ancilla
registers, used to store the potential energy and the intermediate
calculation results. If we assume a simulation of a $d$-dimensional
system in which each Cartesian coordinate is divided into a uniform
grid of $N=2^{n}$ points, the representation of the wavefunction
requires a total of $nd$ qubits in $d$ registers. As for the ancilla
registers, their total number will depend on the complexity of
evaluating the potential and the kinetic energy. At least one
register is always required, to be used as the target of addition
for the purpose of phase kickback. The ancilla registers will
require $m$ qubits each, where $m$ is
chosen in such a way that the registers can store the value of
$V(x)$ with desired accuracy, in the form of a binary integer
between $0$ and $2^m$.

The time required for the simulation is the number of elementary
(one- and two-qubit) gates required to perform the algorithm. Except
in trivial cases, the evaluation of the potential energy will be
much more complicated than that of the kinetic energy $T$, which is
a simple quadratic form. We therefore approximate the total gate
count as being equal to the complexity of evaluating the potential:
even for the simple Coulomb potential, the error thus
introduced to the resource count is substantially less than one percent.

\textbf{Coulomb potential.} The simulation of chemical dynamics depends on
computing the Coulomb potential, and here we provide a detailed
count of the resources required for evaluating it on a quantum
computer. We begin by developing some necessary quantum arithmetic.

For addition, we adopt Draper's quantum addition algorithm [1],
which is based on the quantum Fourier
transform (QFT), and requires only $\frac32m^2$ controlled
rotations. While it is not asymptotically optimal
(i.e., it scales as $O(m^2)$ and not $O(m)$ as does the schoolbook
addition algorithm), it both has a small
prefactor that makes it attractive for the addition of small numbers,
and it is easily adapted for multiplication. Subtraction requires the
same number of rotations, except that they are performed in the
opposite direction.

We perform multiplication using the schoolbook method. The first
multiplicand is repeatedly bit shifted and added to the accumulator if
the corresponding bit of the second multiplicand is 1. Since each
number has $m$ bits, we need to make a total of $m$ such controlled
additions (C-ADD). The product will have $2m$ bits, but we will only
keep the $m$ most significant ones, essentially performing
floating-point arithmetic. For the C-ADD, we first apply a QFT to the
accumulator, as in Draper's algorithm (we will also apply an inverse
QFT at the end, and these two require $n^2$ steps in all). Each C-ADD
requires $\frac12m^2$ CC-rotations, which each can be implemented
using two CNOTs and three C-rotations. Hence, each C-ADD requires $\frac52m^2$
operations, giving a total of $\left( \frac52m^3+m^2\right)$ gates for a
multiplication. However, since half of the CC-rotations are to the
insignificant bits of the accumulator and are subsequently discarded,
we only need to perform $\left( \frac54m^3+m^2\right)$ gates for a
multiplication.

To compute the Coulomb potential, the distance $r$ between two
particles,
$r^{2}=(x_{2}-x_{1})^{2}+(y_{2}-y_{1})^{2}+(z_{2}-z_{1})^{2}$, must be
known. Evaluating $r^2$ requires 3 subtractions and 3 squarings (the
two additions are performed automatically since the squarings are
really additions that can use the same accumulator). For squaring, we
multiply the number by itself using the multiplication circuit, giving
the total requirement of $\left( \frac{15}{4}m^3+\frac{15}{2}m^2\right) $ gates
for
computing $r^2$. The same computation would be used in momentum space
for computing $p^2$ or for simulating a harmonic oscillator potential.

The evaluation of the Coulomb potential is complicated by the need
for a square root. Since evaluating $\sqrt{S}$ is just as difficult
as evaluating $1/\sqrt{S}$, we can find $1/r$ from $r^2$ in one
computation using the Newton-Raphson method, with the iteration
$x_{n+1}=\frac{x_{n}}{2}\left(3-r^2\cdot x_{n}^{2}\right)$. The number
of iterations will depend on the desired final accuracy, but
numerical experiments show that for many ranges of $S$, four iterations suffice
to compute $1/\sqrt{S}$ to within less than $0.03\%$ over the entire range.
Each iteration requires one
subtraction and three multiplications (one of them bit-shifted due
to the factor of $\frac{1}{2}$). So the requirement for $1/\sqrt{S}$
is $(15m^3+18m^2)$ gates, which together with calculating the
distance $r^{2}$ gives the total requirement for the Coulomb
potential as $\left( \frac{75}{4}m^3+\frac{51}{2}m^2\right) $ gates for each
pair
of particles.

\textbf{Potentials fitted from first-principles calculations.}
When using the Born-Oppenheimer approximation, one uses a potential
$V(\mathbf{x})$ which is a function of only the nuclear
coordinates. It is the total energy of the molecule
assuming that the electrons are in their ground state given the
potential induced by the nuclei at coordinates $\mathbf{x}$. In
general, this ground state energy is difficult to compute on a
classical computer. Thus, interpolation schemes may be used to
approximate $V(\mathbf{x})$. Here we analyze the computational
resources needed for such schemes.

We represent the potential as a $d$-dimensional interpolating
polynomial:
\begin{eqnarray*}
V(\mathbf{x}) & = & \sum_{k_{1},k_{2}\ldots k_{d}=0}^{K}c_{k_{1}k_{2}\cdots
k_{d}}\, x_{1}^{k_{1}}x_{2}^{k_{2}}\cdots x_{d}^{k_{d}}\\
 & = &
\sum_{k_{1}=0}^{K}b_{k_{1}}x_{1}^{k_{1}}\sum_{k_{2}=0}^{K}b_{k_{1}k_{2}}x_{2}^{
k_{2}}\cdots\sum_{k_{d}=0}^{K}b_{k_{1}k_{2}\cdots k_{d}}x_{d}^{k_{d}}
\end{eqnarray*}
which can be evaluated using Horner's method starting with the
innermost sum. That is, one has to evaluate one order-$K$ polynomial
in $x_{1}$, $K$ such polynomials in $x_{2}$, and so on until
$K^{d-1}$ polynomials in $x_{d}$. That is, the total number of
polynomials that need to be evaluated is
$\sum_{i=1}^{d-1}K^{i}\approx \frac{K^{d}}{K-1}$. In order that the
results of these calculations be available as constants for
higher-level polynomials, all the intermediate polynomial
evaluations have to be saved in temporary memory along the way. The
number of required registers is  $\frac{K^{d-1}}{K-1}$.

Each polynomial that is evaluated has the form
\begin{equation*}
P(x)=\sum_{k=0}^{K}p_{k}x^{k}=p_{0}+x\left(p_{1}\cdots+x\left(p_{k-2}+x\left(p_{
k-1}+xp_{k}\right)\right)\right)
\end{equation*}
where $p_{k}$ is some constant. Therefore, each polynomial
evaluation requires precisely $K$ additions and $K$ multiplications.
As we have seen above, an addition requires $\frac32m^2$ operations,
and a multiplication $\left(\frac54m^3+m^2\right)$, meaning that in total each
polynomial evaluation requires $K\left(\frac54m^3+\frac52m^2\right)$ gates.
Therefore, the total number of gates required to calculate $V$ is
$\frac{K^{d+1}}{K-1}\left(\frac54m^3+\frac52m^2\right)\approx K^d\left(\frac54m^3+\frac52m^2\right)$.
Furthermore, the total qubit requirement is
$n_{\mathrm{tot}}=nd+m\frac{K^{d-1}}{K-1}$.

\begin{small}
\begin{enumerate}
\item T.~G. Draper, quant-ph/0008033 (2000).
\end{enumerate}
\end{small}

\end{document}